\documentclass[referee]{raa}            

\usepackage{graphicx,times}             
\usepackage{natbib}
\usepackage{amssymb,amsmath}
\bibpunct{(}{)}{;}{a}{}{,}

\usepackage[papersize={8.5in,11in}]{geometry}

\begin{document}

\title{Evolution of close binary systems parameter distributions}

   \volnopage{Vol.0 (20xx) No.0, 000--000}      
   \setcounter{page}{1}          

   \author{Dugasa Belay Zeleke\inst{1}, Seblu Humne Negu\inst{1}, Oleg Yu. Malkov\inst{2}}
   

   \institute{Ethiopian Space Science and Technology Institute (ESSTI),
              Entoto Observatory and Research Center (EORC),
              Astronomy and Astrophysics Research and Development Department,
              P.~O.~Box 33679 Addis Ababa, Ethiopia. {\it dugasa32@gmail.com}, {\it seblu1557@gmail.com}\\
        \and
             Institute of Astronomy, Russian Academy of Sciences, 48 Pyatnitskaya St., Moscow 119017,
             Russia. {\it malkov@inasan.ru}\\
\vs\no
   {\small Received~~20xx month day; accepted~~20xx~~month day}}

\abstract{In this paper, we investigate the orbital and stellar parameters of
low- and intermediate-mass close binary systems.
We use models, presented in the catalogue of \cite{2000MNRAS.319..215H} and calculate
parameters of accretors.
We also construct distributions of systems along luminosity, semi-major axis and angular momentum,
and make some conclusions on their evolution with time.
 We made a comparison of the results with observational data and it shows a good agreement.
The set of theoretical models published in \cite{2000MNRAS.319..215H} quite adequately
describes the observational data
and, consequently, can be used to determine the evolutionary path of specific close binary systems,
their initial parameters values and final stages.
}

   \authorrunning{D. B. Zeleke, S. H. Negu \& O. Yu. Malkov}            
   \titlerunning{Close binaries parameter distributions}  
   \maketitle

%
\section{Introduction}           
\label{sect:intro}
The evolution of binary stars is similar to that of single stars unless there is some form of mass transfer between the 
two stars. Individual stars are not affected by the presence of a companion if the binary orbit is wide enough, so 
standard stellar evolution theory is all that is required to describe their evolution. However, if the stars get close 
enough, they can interact, with consequences for the stars evolution and orbit. Close binary evolution has been 
observed 
in many systems, including cataclysmic variables, X-ray binaries, and Algols, as well as in the presence of stars such 
as blue stragglers, which cannot be explained by single star evolution. 

The majority of stars are binary or multiple systems \citep{duquennoy1991multiplicity, raghavan2010survey}. Those stars 
in interacting binary systems exchange mass and angular momentum through material transfer between the two stars, and 
the systems as a whole may lose both mass and angular momentum when the material is ejected from the systems.

In binary systems mass transfer will proceed with different stellar evolution models \citep{shore1994observations}. In 
a 
close binary system, mass transfer usually occurs via Roche lobe overflow (RLOF), in which the primary star begins to 
transfer material to its companion star once it fills its Roche lobe \cite{kopal1959close, paczynski1971evolutionary, 
eggleton1983approximations} and if one of the stars ejects much of its mass in the form of stellar wind. The mass 
transfer from one star to another causes the angular momentum and orbital period of these systems. There are two 
mechanisms of mass transfer between the components of close binaries. These are conservative mass transfer, in which 
both the binary's mass and angular momentum are conserved, and nonconservative mass transfer, in which both the 
binary's 
mass and angular momentum decay with time. As a result, there are observational evidences for both conservative and 
nonconservative mass transfer in close binaries \cite{podsiadlowski2001evolution, yakut2006observational, 
manzoori2011mass, pols2012lecture}.  More recent discussions on observational proves of conservative and 
non-conservative scenarios of mass transfer can be found in \cite{2020RAA....20..163Q, 2018MNRAS.481.5660D, 
2021arXiv211114047P, 2022JPhCS2214a2005V, 2022MNRAS.514..622M}, while these and other problems related to the evolution 
of close bianries are reviewed in \cite{2020PhyU...63..209T, 2022arXiv220310066O}.

If one star loses mass via a stellar wind, the companion may accrete some of the material, and affecting the orbit. In 
these systems, the tidal interaction plays an important role in changing the orbit of a close binary system. The degree 
of interaction is critically dependent on the stellar radius to the star separation ratio \cite{zahn1977tidal, 
hut1981tidal}. As the binary approaches an equilibrium state of minimum energy, tide can synchronize the spin of the 
stars with the orbit and circularize an eccentric orbit. The existence of a companion introduces a tidal, which acts to 
elongate the star along the line between the centers of mass, resulting in tidal bulges. As noted by 
\cite{hurley2002evolution} if the rotational period of the stars is shorter than the orbital period, frictional forces 
on the star's surface will drag the bulge axis ahead of the line of centres.

In this paper, we concentrate on determining the angular momentum and orbital period evolution of close binary systems 
at various events, such as the starting of RLOF, the minimum luminosity during RLOF and the end of the last episode of 
RLOF from the catalogs of \cite {2000MNRAS.319..215H}. Then, in order to determine the evolution of close binary 
systems, we examine the orbital and stellar parameters of close binary systems such as semi-major axis, mass, 
luminosity, effective temperature, and radius of the accretor stars in the range of masses of the donor between $0.212 
M_{\odot}$ and $7.943 M_{\odot}$ and mass of accretor between $0.25 M_{\odot}$ to $14.281 M_{\odot}$.

The goal of this paper is to determine the angular momentum and orbital period evolution of close binary stars from 
catalog of \cite{ 2000MNRAS.319..215H}. We will look at the angular momentum, semi-major axis, mass, luminosity, 
effective temperature, and radius of accretor stars. Finally, we present the statistical analysis as well as the 
comparison with observations. The paper is structured as follows. The basic mathematical formulations for orbital and 
stellar parameters of these systems are presented in Section (\ref{sec:evol}), and the analysis of data is presented in 
Section (\ref{sec:analy}). Finally, in Section (\ref{sec:conc}) we make our conclusions

\section{EVOLUTIONS OF CLOSE BINARY SYSTEMS}\label{sec:evol}
\subsection{Basic Assumptions}
In this paper, we look at close binary systems with accretor masses $M_a$, donor masses $ M_d$,
and the total mass of star which is, $M_t = M_ a + M_ d$,
and volume equivalent radii of $R_a$ and $R_d$.
We assume that each star's mass is distributed spherically symmetrically.
The binary is assumed to be in a circular Keplerian orbit with semi-major axis $a$,
orbital angular velocity $\Omega_{orb}$
and orbital period, $P_{orb}$, at initial phase. 

\subsection{Angular momentum evolution}
The matter is transferred between the components in the binary systems,
and the orbital period is changed as a result of the angular momentum redistribution 
between the two stars.
Orbital angular momentum of the binary system with a circular orbit is given by 

\begin{equation}
 J_{orb}= M_a M_d\sqrt\frac{Ga}{M_t} \label{eqn:1}
\end{equation}
where $a$ is semi-major axis, $G$ is universal gravitational constant and $M_t$ is the total mass of the stars.

The total mass of the stars in terms of the mass of accretor and mass ratio can be written as
\begin{equation}
 M_t= M_a(1+q) \label {eqn:2}
\end{equation}
where $q= \frac{M_d}{M_a}$, which is the mass ratio of the system.

Substituting Eq.~\eqref{eqn:2} into Eq.~\eqref{eqn:1} the orbital angular momentum can be given as
\begin{equation}
 J_{orb} = M_a M_d\left(\frac{Ga}{M_a(1+ q)}\right)^\frac{1}{2} \label{eqn:3}
\end{equation}
Hence, the semi-major axis of these systems will be obtained by:
\begin{equation}
 a= \left(\frac{GP^2_{orb} (M_a(1+q)}{4\pi^2}\right)^\frac{1}{3}. \label{eqn:4}
\end{equation}

As noted by \cite {negu2015mass}, the angular momentum stored in the rotation of the two stars is negligible in comparison to the orbital angular momentum, 
so Eq.~\eqref{eqn:3} approximates the angular momentum to the binary.
We obtain a general expression for orbital evolution by differentiating Eq.~\eqref{eqn:3}:
 
\begin{equation}
 2\frac{\dot J_{orb}}{J_{orb}}= \frac{\dot a}{a} + 2\frac{\dot M_a}{M_a} +2\frac{\dot M_d}{M_d} -\frac{\dot M_a + \dot M_d}{M_a(1+q)},
\label{eqn:5}
\end{equation}
where
$\dot J_{orb}$ denotes angular momentum loss from the binary, which can be caused by gravitational wave radiation 
or mass loss from the binary as a whole or from the component stars.
 
The total mass and angular momentum of the binary systems are conserved in the case of conservative mass transfer.
Hence, we can set $\dot J_{orb}=0$ and $\dot M_a=-\dot M_d$.
Then, Eq.~\eqref{eqn:5} reduced to:
\begin{equation}
 \frac{\dot a}{a}= 2\left(\frac{M_d}{M_a}-1\right)\frac{\dot M_d}{M_d} , \label{eqn:6}
\end{equation}
Eq.~\eqref{eqn:6} tells us that, when $\dot M_d<0$, the orbit shrinks $(\dot a<0)$ as long as $ M_d>M_a$
the orbit and expands when $M_d<M_a$ \citep{paczynski1971evolutionary}.

\subsection {Stellar parameters relations for close binary stars}

The relationship between a star's mass and luminosity is a fundamental law that is used in many fields of astrophysics.
It is particularly important in the construction of the initial mass function from the luminosity function of stars \citep{malkov1997mass}.

In these systems, the mass-luminosity relationship of the stars is determined by
\begin{equation}
 L \propto M^n  \label{eqn:16}
\end{equation}
with a different range of n that can be used to determine the stage of close binary stars, which is dependent on the 
mass of the stars.

In accordance with \cite {duric2004advanced}, the accretor star's luminosity can be calculated with
\begin{equation}
 L_a= L_{\odot}\left(\frac{M_a}{M_{\odot}}\right)^4  \label{equ:accL}
\end{equation}

The mass-radius relation can also be expressed over a wide range of stellar masses as:
\begin{equation}
  R\propto  M^n \label{eqn:20}
\end{equation}
with various n ranges that can be used to determine the stage of close binary stars.
As noted by \cite{bonnell2005binary}, we can express the accretor star's radius in terms of solar mass and radius:
\begin{equation}
 R_a= R_{\odot}\left(\frac{M_a}{M_{\odot}}\right)^{0.8} \label{equ:accR}
\end{equation}
A comparison of theoretical and empirical stellar mass-radius relations can be found
in \cite{malkov2007mass, eker2015main}.

The star's luminosity can be also expressed as follows:
\begin{equation}
  L=4\pi R^2\sigma T^4_{eff} \label{eqn:17}
\end{equation}
where $\sigma$ denotes the Stefan-Boltzmann constant and $T_{\rm eff}$ denotes
the effective temperature of the stars.
Consequently, using Eq.~\eqref{eqn:17} the effective temperature of the accretor star can be obtained by:
\begin{equation}
 T_a=\left(\frac{L_a}{4\pi R^2_a\sigma}\right)^\frac{1}{4} \label{equ:accT}.
\end{equation}

\section{Analysis of data for low and intermediate-mass close binaries}\label{sec:analy}
\label{parameters}

\subsection{Stellar parameters and comparison with observations}

Here we analyse data for donors and accretors of
low and intermediate-mass close binaries from catalogue of
\cite{2000MNRAS.319..215H} and compare them with observational data.
Astrophysical parameters of donors are given in \cite{2000MNRAS.319..215H},
accretors' masses were calculated from donors' masses and mass ratio $q$.
Luminosity and radius of accretors were estimated with Eqs. \eqref{equ:accL}
and \eqref{equ:accR}, respectively, and temperature was calculated from
\eqref{equ:accT}.

We study results of calculations made by \cite{2000MNRAS.319..215H} for
three events, namely, (a) the beginning of RLOF, (b)
minimum luminosity during RLOF, and (c) the end of the last episode of RLOF,
hereafter a-event, b-event and c-event.

\begin{figure}
\centering
\includegraphics[width=7cm]{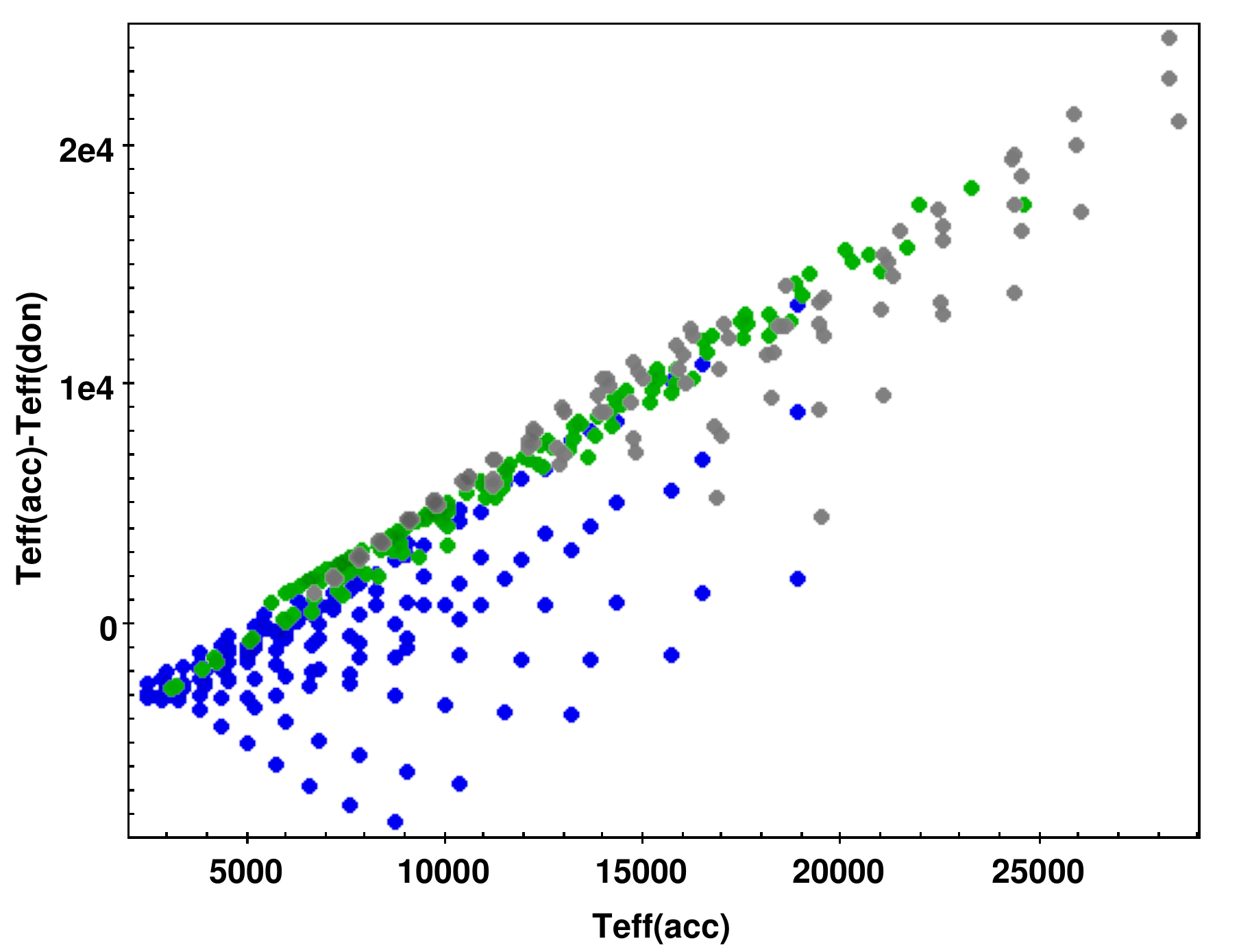}
\includegraphics[width=7cm]{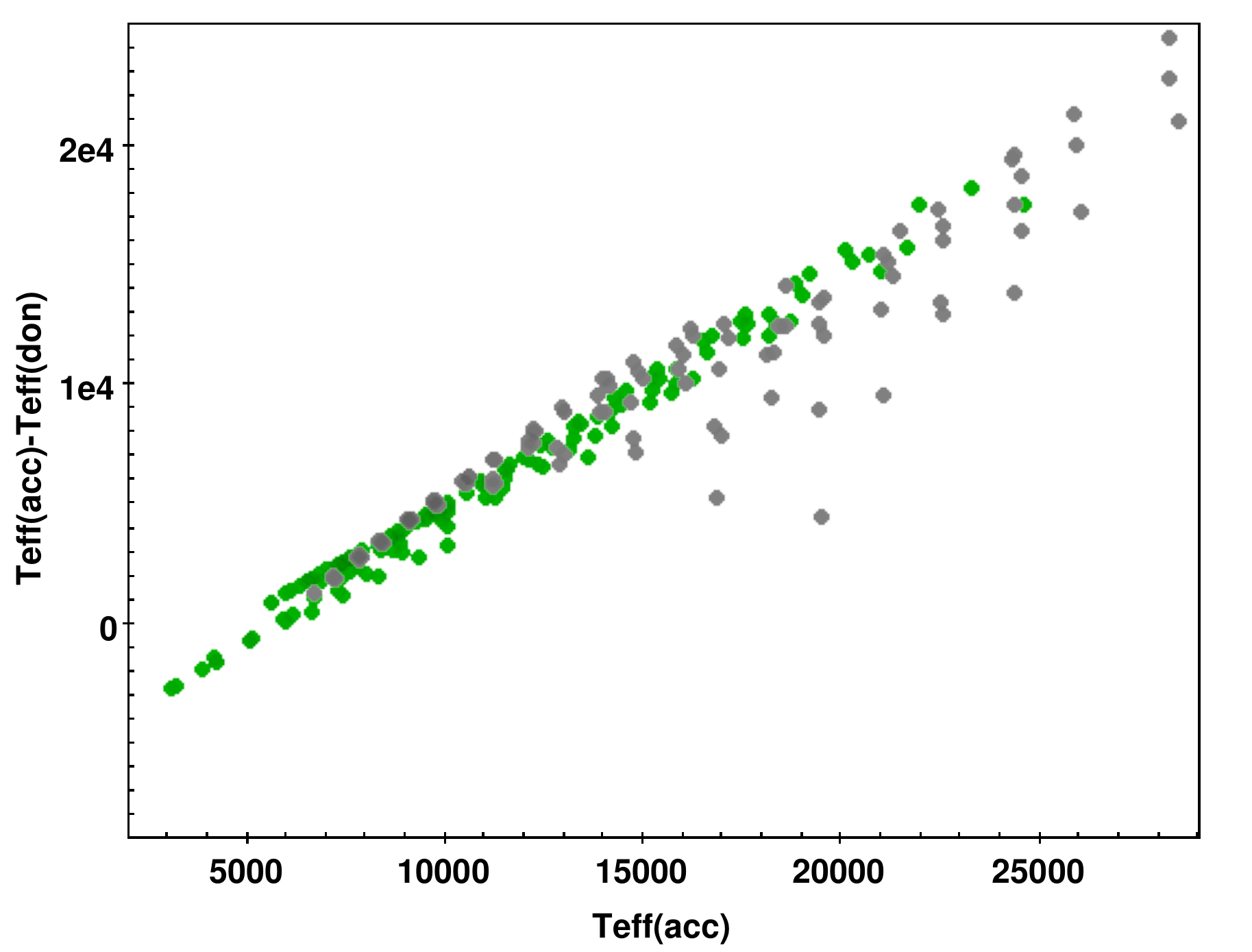}
\caption{Effective temperatures of accretors and donors of systems from \cite{2000MNRAS.319..215H}.
Blue, green and grey colors represent stars at the beginning of RLOF (a-event),
at minimum luminosity during RLOF (b-event), and at the end of the last episode of RLOF (c-event).
Left panel: all three events, right panel: all events except the first.}
\label{fig:Teff}
\end{figure}

First, analyzing $T_{\rm eff}$ of accretors and donors (Fig.~\ref{fig:Teff}),
we should note that, except for a-events,
the ``temperature -- temperature'' relation for cool stars $T_{\rm eff}<14000$ K
can be satisfactory approximated by
\begin{equation}
T_{\rm eff}({\rm acc})-T_{\rm eff}({\rm don}) = 0.99 \times T_{\rm eff}({\rm acc}) - 5260 K
\end{equation}
This relation is in an excellent agreement with the relation found by \cite{2021A&AT...32..111M}
for 119 systems, included in the comprehensive list of semi-detached double-lined
eclipsing binaries:
\begin{equation}
T_{\rm eff}({\rm acc})-T_{\rm eff}({\rm don}) = 0.9 \times T_{\rm eff}({\rm acc}) - 4000 K
\end{equation}
Stars in the beginning of RLOF (a-event) do not satisfy this relation.
This is not surprising, as this stage is very short-lived and therefore is extremely rarely observed.
Consequently, such stars are not included in the catalogue of \cite{2021A&AT...32..111M}.

\begin{figure}
\centering
\includegraphics[width=7cm]{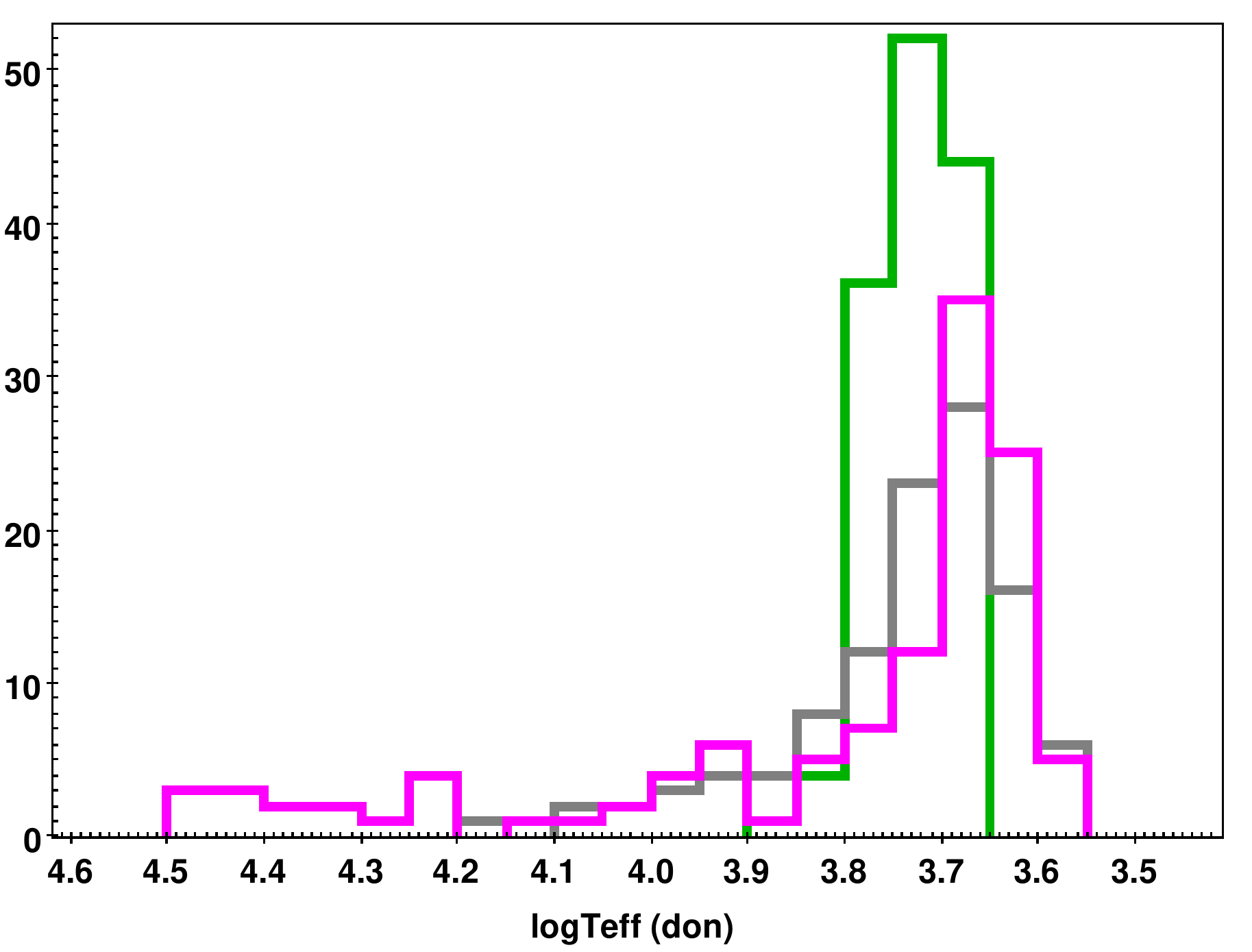}
\includegraphics[width=7cm]{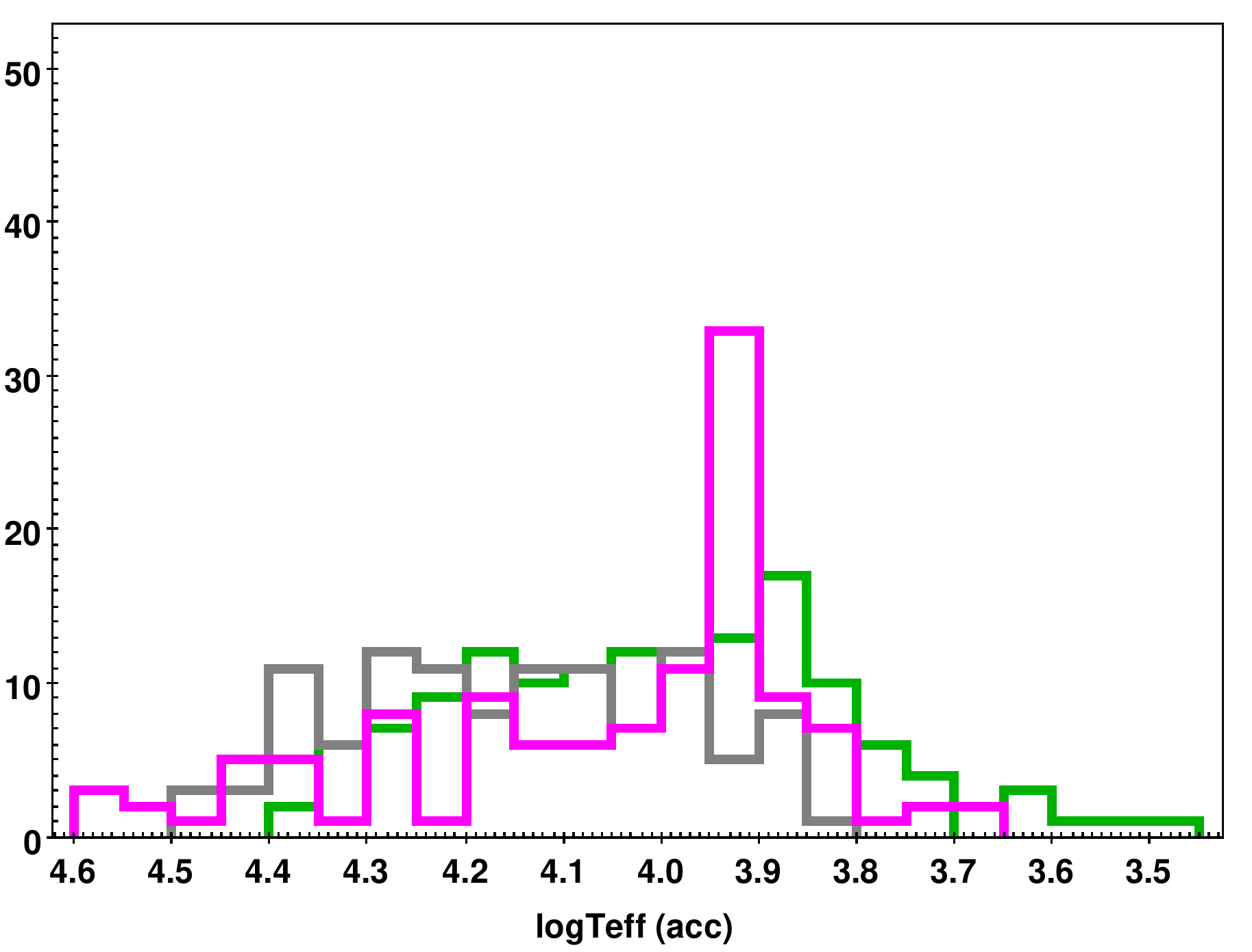}
\caption{Effective temperatures of donors (left) and accretors (right) of systems from \cite{2000MNRAS.319..215H}.
Green and grey colors represent stars at minimum luminosity during RLOF (b-event), and at the end of the last episode 
of RLOF (c-event), respectively. Observational data for semi-detached double-lined eclipsing binaries from 
\cite{2020MNRAS.491.5489M} are shown as pink histograms. Note that the X-axes are flipped.}
\label{fig:hisTeff}
\end{figure}

Distribution of $T_{\rm eff}$ for donors (left) and accretors (right) is shown in Fig.~\ref{fig:hisTeff}.
These distributions demonstrate a good agreement with ones constructed from observational data
for semi-detached double-lined eclipsing binaries \cite{2020MNRAS.491.5489M},
 shown as pink histograms in Fig.~\ref{fig:hisTeff}.
The time it takes for a close binary star to go from a-event to b-event is quite short,
so almost all known semi-detached systems are observed between b-event and c-event
(see also Fig.~\ref{fig:q}, right panel). That is why the observational distributions
are qualitatively consistent with theoretical ones, and
the positions of the maximum for observed temperatures are close to the theoretical ones.

\begin{figure}
\centering
\includegraphics[width=7cm]{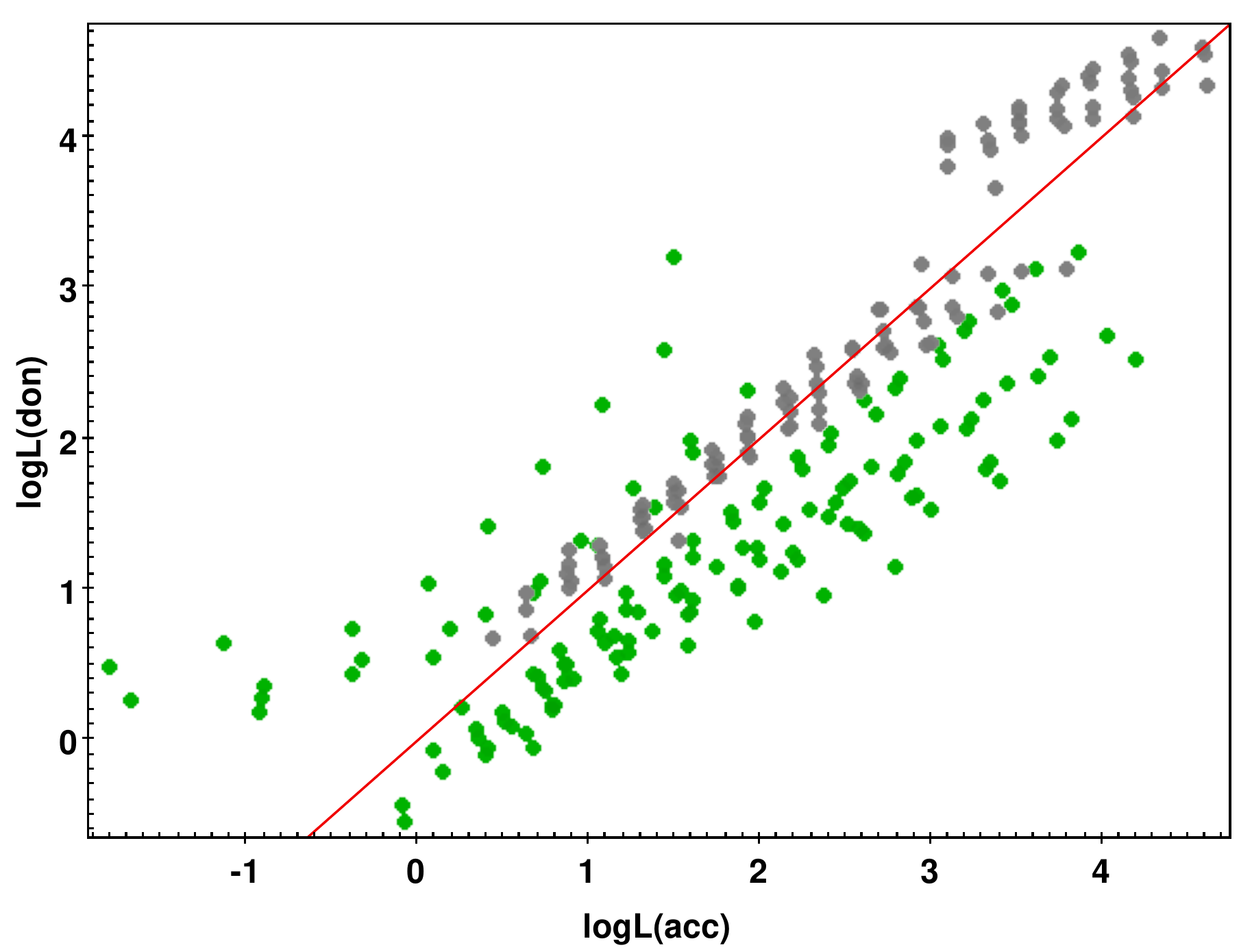}
\caption{Luminosities of donors and accretors of systems from \cite{2000MNRAS.319..215H}.
Colors are as in Fig.~\ref{fig:hisTeff}.
The one-to-one relation is also shown as a thin red line, for reference.}
\label{fig:L}
\end{figure}

\begin{figure}
\centering
\includegraphics[width=6cm]{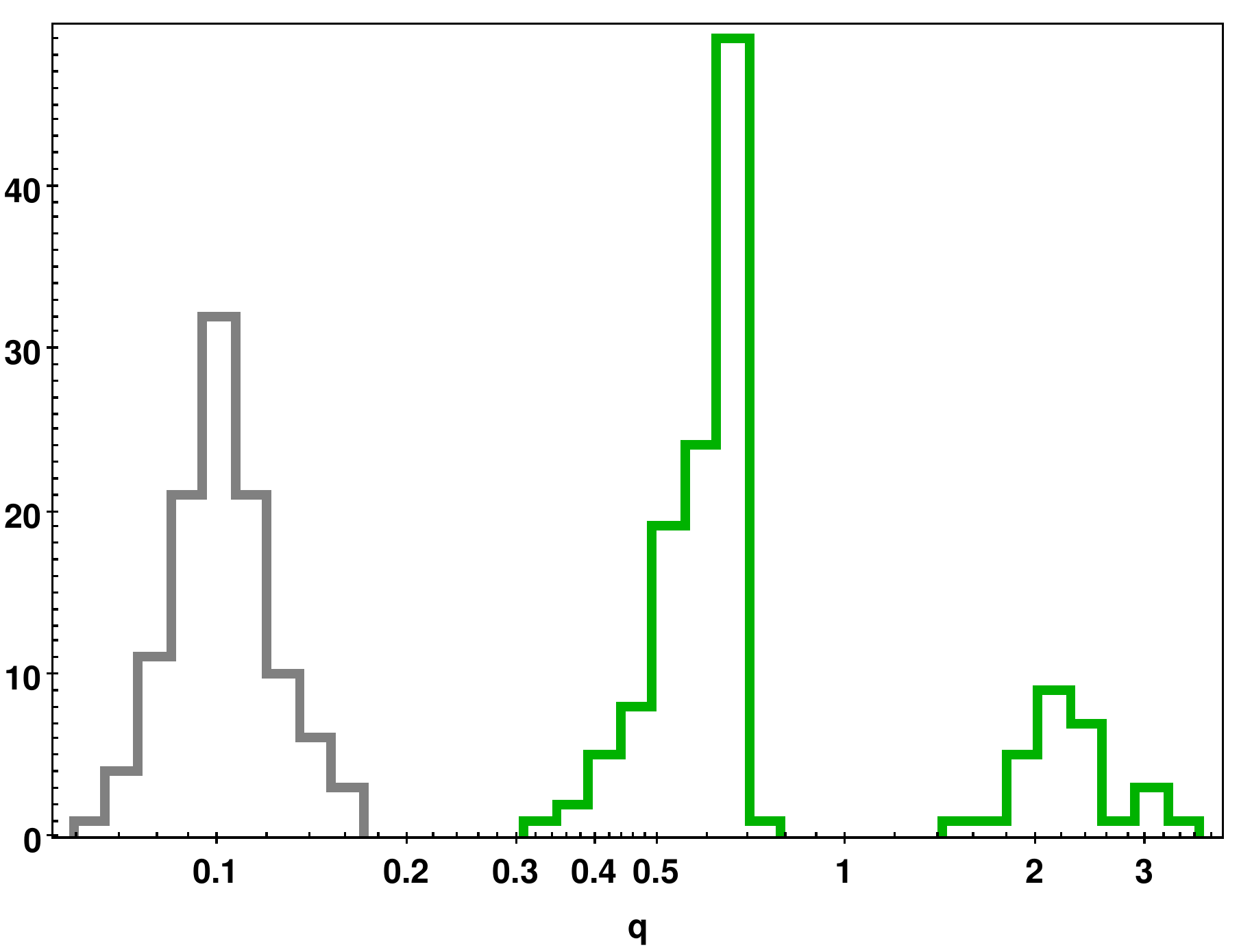}
\includegraphics[width=6cm]{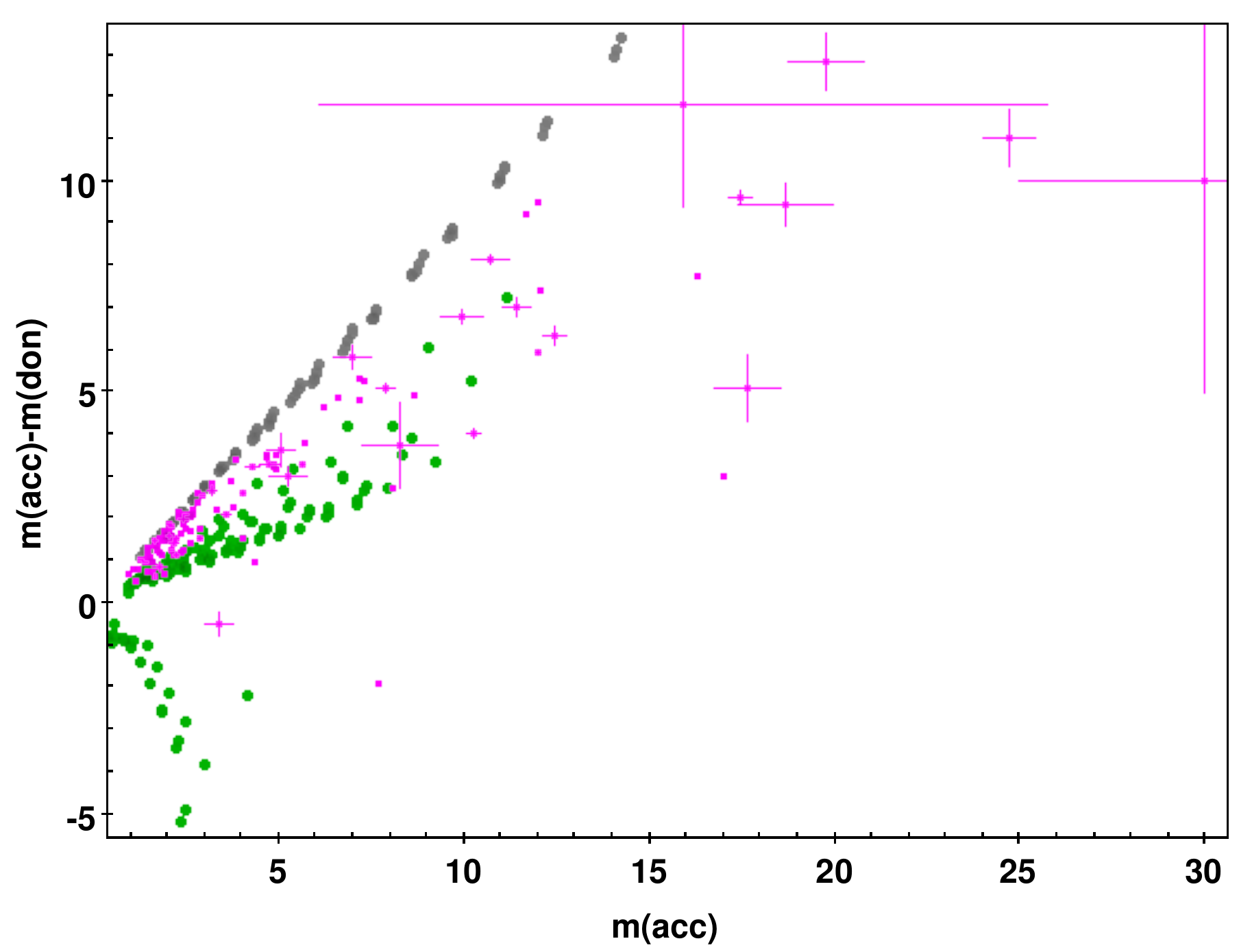}
\caption{Mass ratio $q=M_{don}/M_{acc}$ and masses of systems from \cite{2000MNRAS.319..215H}.
Green and grey colors represent stars at minimum luminosity during RLOF (b-event), and at the end of the last episode 
of RLOF (c-event), respectively. Left panel: distribution of $q$ (note a logarithmic scale for q-axis), right panel: 
donor-accretor mass relation.
 Observational data for semi-detached double-lined eclipsing binaries from \cite{2021A&AT...32..111M}
are shown as pink dots, with observational errors.
}
\label{fig:q}
\end{figure}

Fig.~\ref{fig:L} demonstrates luminosities of donors and accretors for systems from \cite{2000MNRAS.319..215H}.
An agreement with observational data (see Fig.~5 in \cite{2020MNRAS.491.5489M}) is less satisfactory, as the
majority of observational points lie on or below the $L_{\rm don}=L_{\rm acc}$ line.
The paucity of observational data above that line, however, can be explained.
A grey point group in the upper right represent the most luminous (i.e., most massive) stars that are
extremely rare in the solar neighbourhood. Green points are pairs with relatively large mass ratio $q$, and,
consequently, large magnitude difference prevents such stars from being detected as spectroscopic binaries.

It is interesting also to look at the mass ratio $q=M_{don}/M_{acc}$ distribution of systems,
included in the catalogue of \cite{2000MNRAS.319..215H} (Fig.~\ref{fig:q}, left panel).
Systems at the end of the last episode of RLOF (c-event)
demonstrate a sharp maximum at $q \sim 0.1$ while distribution of b-event systems
(minimum luminosity during RLOF) is bimodal one (left panel). There is a
sharp maximum at $q \sim 0.7$ and a broader one at $q \sim 2.1$ (note a logarithmic scale for x-axis).

``Large-q'' group comprises initially relatively close systems with low-mass accretor.
During their evolution donors relatively quickly attain minimum luminosity
(the time elapsed between a-event and b-event is minimal for these systems)
so that hydrogen abundance at donor's surface has no time to change from its
initial value $H=0.7$, according to \cite{2000MNRAS.319..215H} data.
Meanwhile their further evolution slows down significantly (c-events for these systems
are not reached in the calculations of \cite{2000MNRAS.319..215H}).

A gap between these two maximums is a consequence of the discreteness of the calculation grid.
\cite{2000MNRAS.319..215H} did the calculations for the initial $q$ values of 1, 1.5, 2, 3, 4.
All pairs with initial $q_{ini}=4$ are located in the $q \sim 2.1$ maximum,
while all pairs with initial $q_{ini}=3$ (with two exceptions) or less are located
in the $q \sim 0.7$ maximum.
 The observation points (Fig.~\ref{fig:q}, right panel)
are located between the b-event and c-event points, because,
as mentioned above, the transition of the star from a-event to b-event is very fleeting.

\subsection{Evolution of stellar parameters}

\begin{figure}
\centering
\includegraphics[width=6cm]{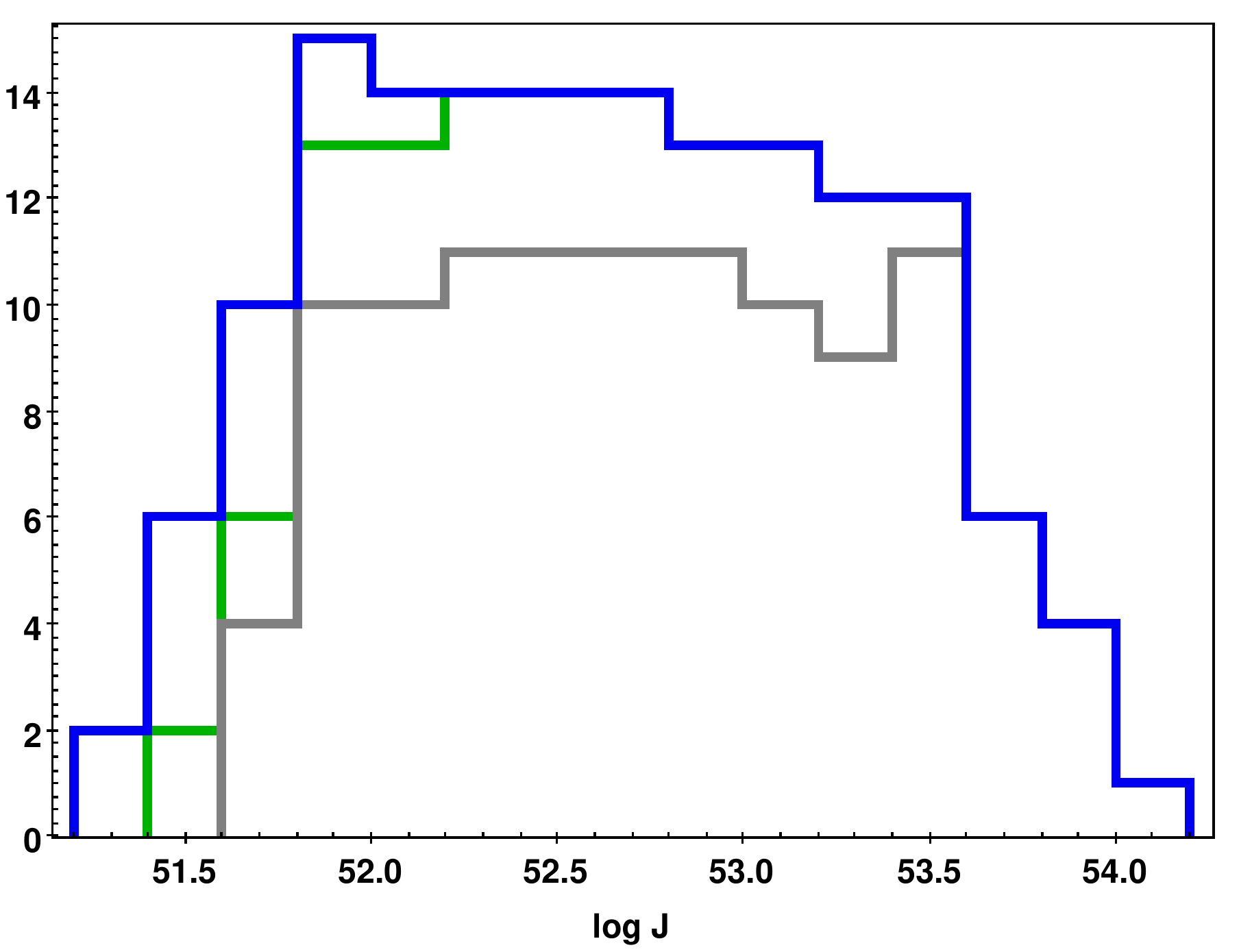}
\includegraphics[width=6cm]{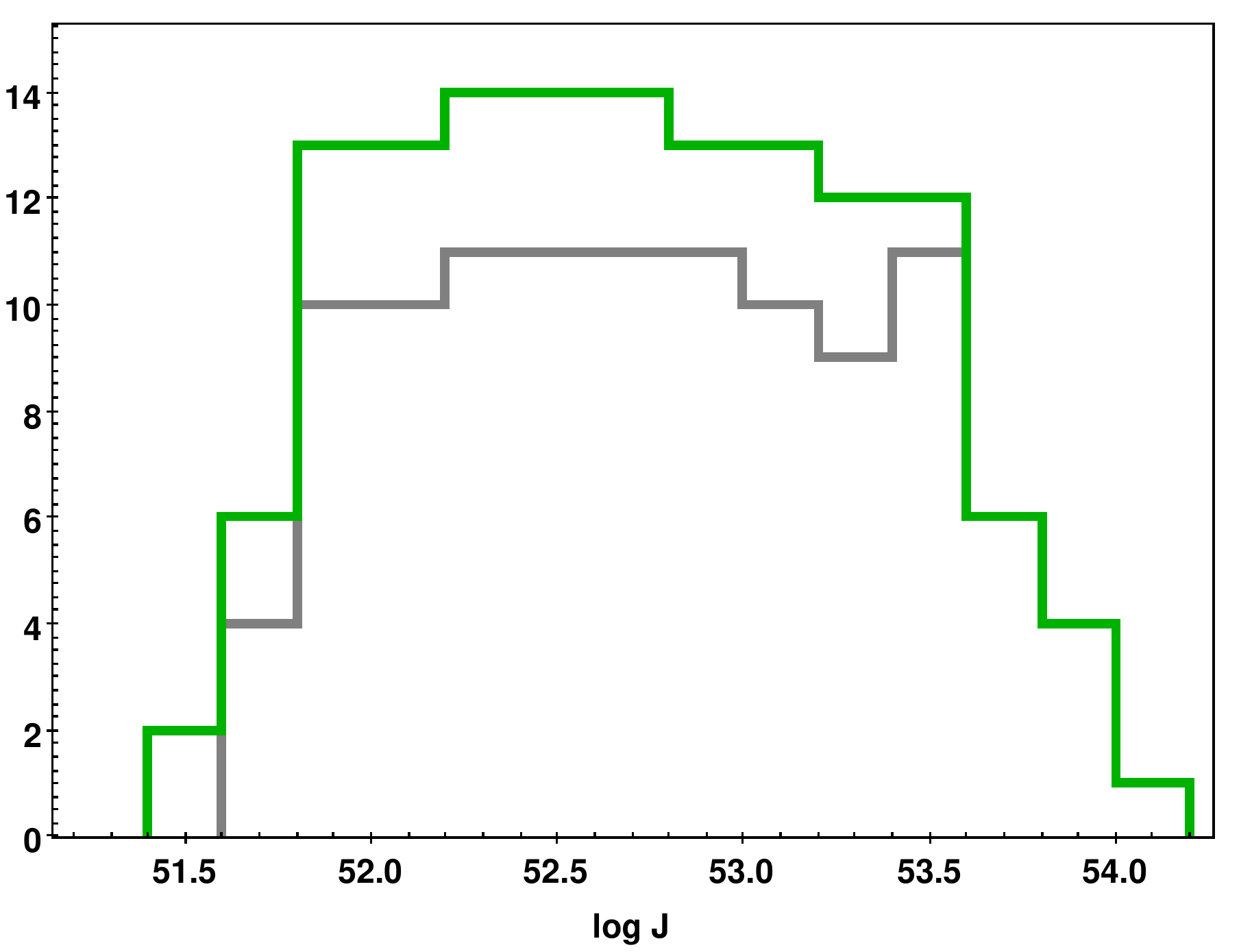}
\caption{Angular momentum of systems from \cite{2000MNRAS.319..215H}.
Blue, green and grey colors represent stars at the beginning of RLOF (a-event),
at minimum luminosity during RLOF (b-event), and at the end of the last episode of RLOF (c-event).
Left panel: all three events, right panel: all events except the first.}
\label{fig:evJ}
\end{figure}

It is advisable to study the behaviour of an ensemble of stars, modeled by \cite{2000MNRAS.319..215H}.
Fig.~\ref{fig:evJ} shows evolution of angular momentum $J$, calculated according to \eqref{eqn:3}.
The b-event and c-event distributions look similar, and they are what we should get from observations,
as they are relatively long evolutionary stages.
Contrary, the ``initial'' (a-event) distribution differs significantly from two others.
 The increase of the angular momentum from the stage when stars are at the beginning of RLOF (a-event)
to the subsequent stages, can be easily explained.
In the process of the binary evolution with mass exchange, the masses of components
become comparable, which, with a fixed sum of masses, increases their product (see Eq.~\eqref{eqn:2}).
It should be taken into account when one restores an initial $J$-distribution from observations.

\begin{figure}
\centering
\includegraphics[width=6cm]{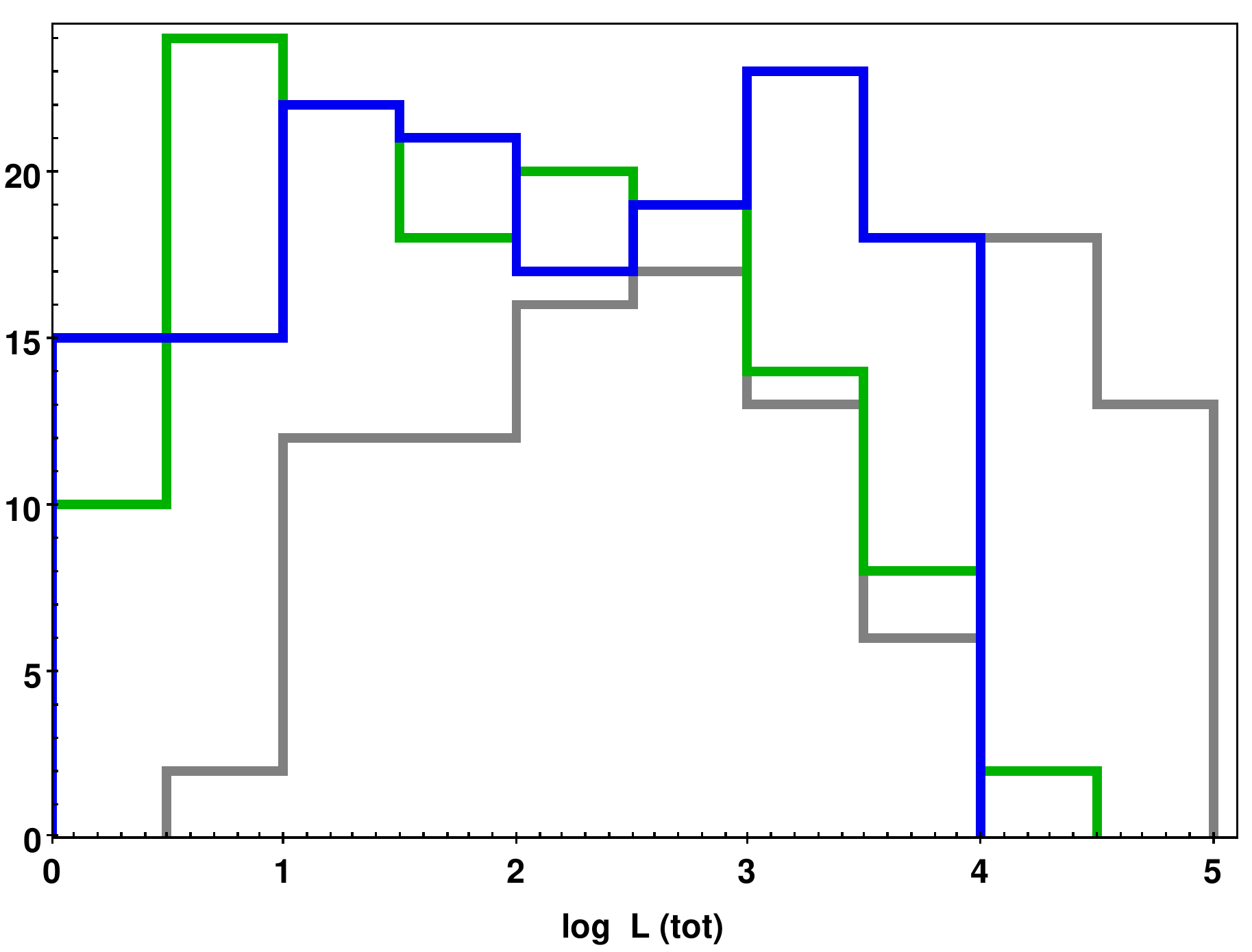}
\includegraphics[width=6cm]{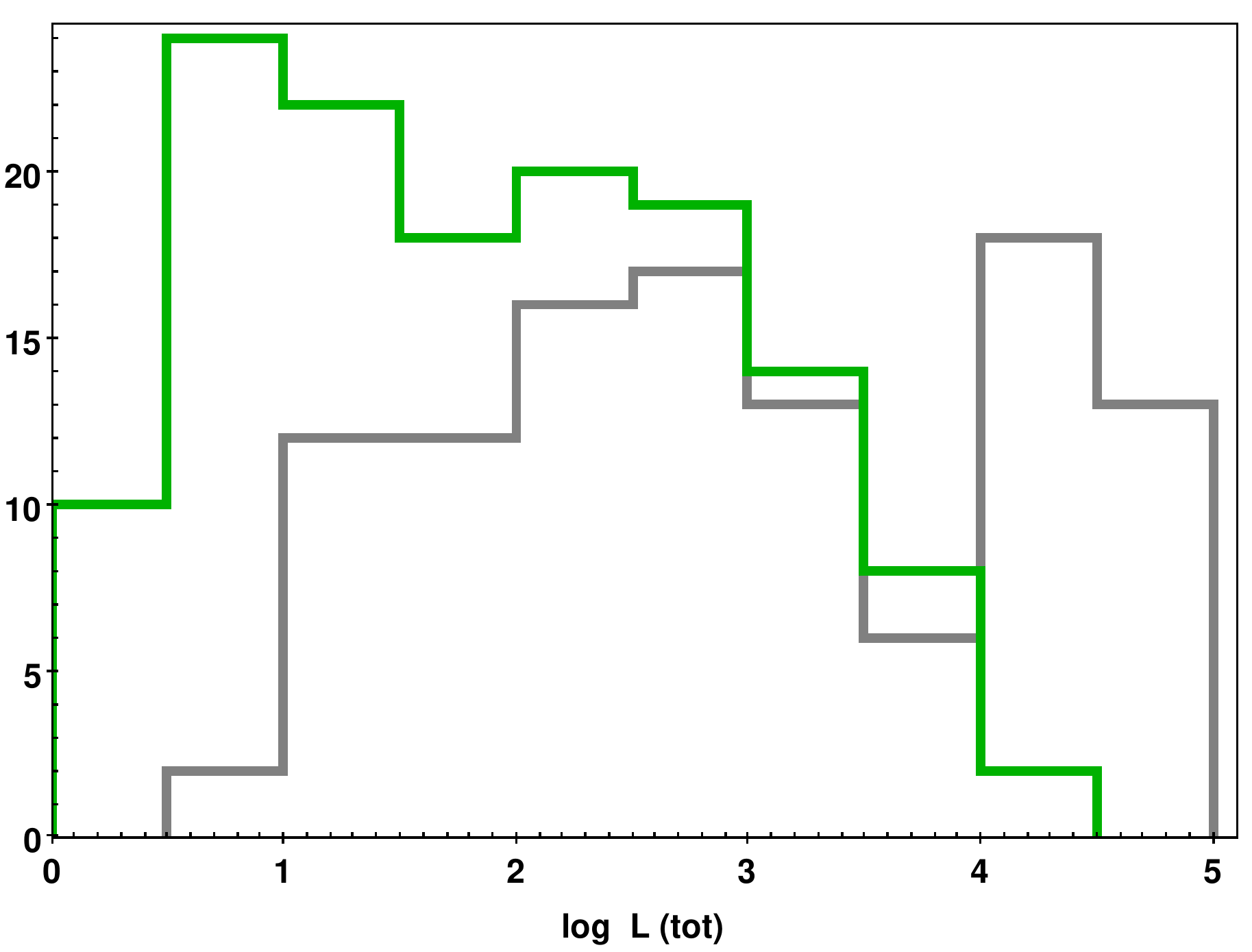}
\caption{Total luminosity of systems from \cite{2000MNRAS.319..215H}.
Colors and panels content are as in Fig.~\ref{fig:evJ}.}
\label{fig:evL}
\end{figure}

Analyzing of evolution of total luminosity of the systems $L_{\rm tot}=L_{\rm don}+L_{\rm acc}$
(see Fig.~\ref{fig:evL}), one can see that distributions for a-event and b-event look similar,
while c-event distribution demonstrates an excess of high luminous systems.
 This is because the b-event, by definition, corresponds to the minimum luminosity,
and at later stages the luminosity of the system begins to increase.
On the one hand, this makes them easier to detect in this evolutionary stage (c-event),
but on the other hand, such systems, being massive, should not be too common in the solar vicinity.

\begin{figure}
\centering
\includegraphics[width=6cm]{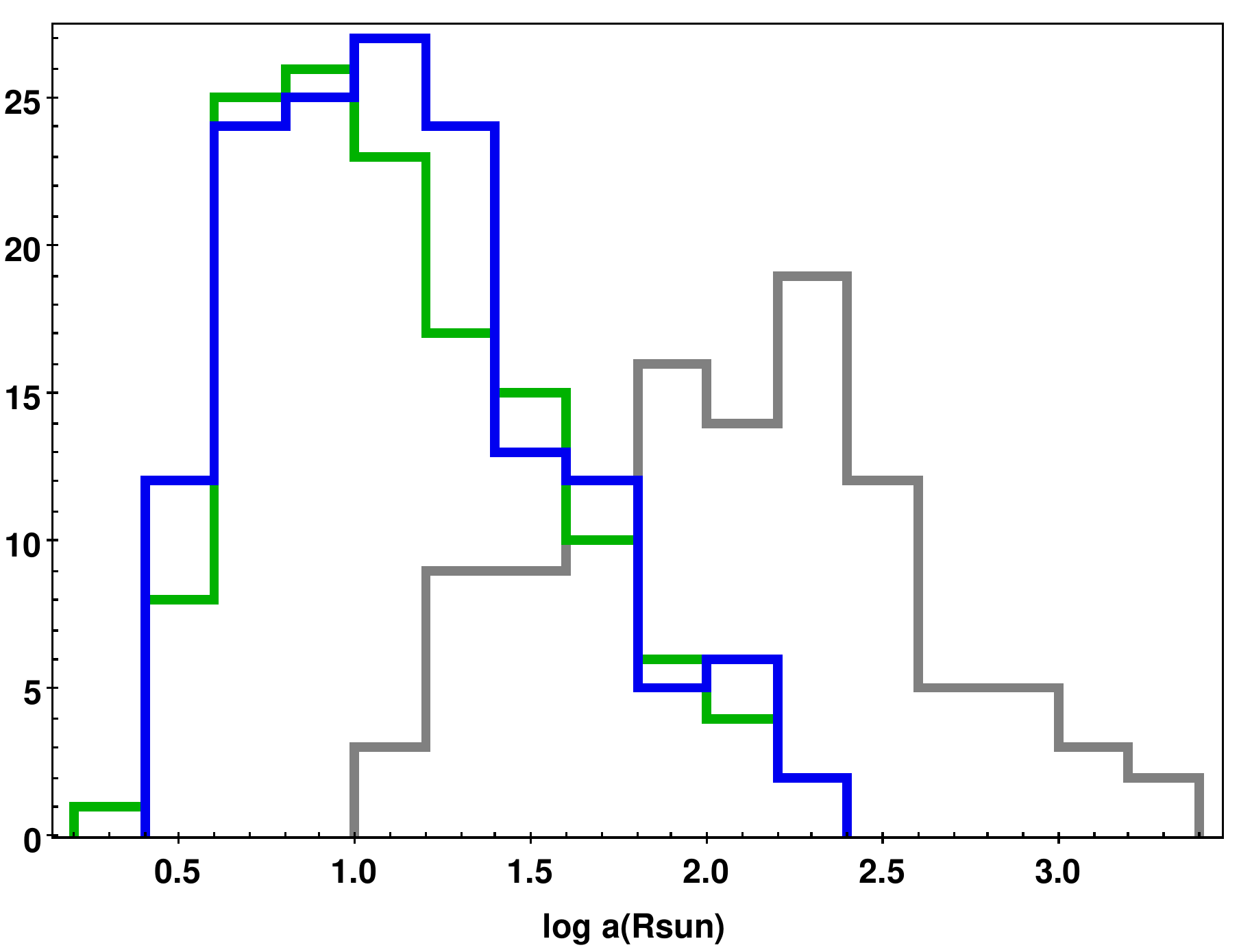}
\caption{Semi-major axis of systems from \cite{2000MNRAS.319..215H}.
Colors are as in Fig.~\ref{fig:evJ}.}
\label{fig:eva}
\end{figure}

\newpage
 Finally, semi-major axis ($a$) distributions (see Fig.~\ref{fig:eva})
demonstrate a quite understandable excess of wide pairs among well evolved c-event systems.
A study of evolution of semi-major axis ($a$) distribution allows predictions to be made
about the frequency of such systems among resolved spectroscopic binaries (RSB),
systems demonstrated spectral lines shift and, at the same time, observed with interferometric techniques.
Minimum $a$ value of a close system to be resolved is about 15 $R_{\odot}$,
so the majority of systems at c-event and some of the systems at the earlier events
can be in principal observed as RSB.
This means that among the systems detected as RSB, there may be pairs just after or even at the mass exchange stage.

\section{Conclusion} \label{sec:conc}
For models of low- and intermediate-mass close binary stars, listed in the catalogue of \cite{2000MNRAS.319..215H}, we 
have estimated parameters of accretors, using catalogued data on donors and assuming that accretors satisfy main 
sequence relations. The results were compared with observational data, collected and discussed in 
\cite{2020MNRAS.491.5489M} and \cite{2021A&AT...32..111M}. We can draw the following conclusions.

 The set of theoretical models published in \cite{2000MNRAS.319..215H} quite adequately (taking into account the 
observational selection effects) describes the observational data and, consequently, can be used to determine the 
evolutionary path of specific close binary systems, their initial parameters values and final stages. Theoretical and 
observational results for effective temperature and mass show a good agreement. Derived relation between effective 
temperatures of donor and accretor can be used for estimation of the former assuming that accretor is still a MS-star.
Results for luminosity show a worse agreement, and we have explained this by selection effects. In general, we have 
shown that observed parameters (in particular, mass ratio $q$) can tell us a lot about origin and current evolutionary 
state of the system.

Study of distributions of the systems on luminosity, semi-major axis and angular momentum, and their evolution with time 
allow us to give some suggestions on the connection between initial and present-day distributions, as well as to make 
some assumptions on a presence of interacted binary stars among resolved spectroscopic binaries.

\begin{acknowledgements}
We are grateful to our reviewer whose constructive comments greatly helped us to improve the paper.
We thank Ethiopian Space Science and Technology Institute, Entoto Observatory and Research Center 
and Astronomy and Astrophysics Research and Development Department for supporting this research.
This research has made use of
SAO/NASA Astrophysics Data System.
\end{acknowledgements}

{99}

\label{lastpage}



\end{document}